%% file: vision.tex
\documentclass[sigconf, nonacm]{acmart}

\usepackage[linesnumbered,ruled,vlined]{algorithm2e}

\usepackage{url}
\usepackage{verbatim}
\usepackage{graphicx}
\usepackage{multirow}
\usepackage{balance}
\usepackage{lscape}
\usepackage{float}
\usepackage{bm}
\usepackage{xcolor}
\usepackage{paralist}
\usepackage{booktabs}
\usepackage{xspace}
\usepackage{listings}
\usepackage{lipsum}
\usepackage{subcaption}
\usepackage{xcolor}
\usepackage{xspace}
\usepackage{subcaption}
\usepackage{caption}
\usepackage{amsmath}
\usepackage{fontawesome}
\usepackage{dsfont}
\usepackage{multirow}

\definecolor{cyanpro}{rgb}{0.0, 0.72, 0.92}
\definecolor{darkterracotta}{rgb}{0.8, 0.31, 0.36}
\definecolor{skyblue}{rgb}{0.53, 0.81, 0.92}
\definecolor{trueblue}{rgb}{0.0, 0.45, 0.81}
\definecolor{blue2}{HTML}{008AFF}
\definecolor{red3}{HTML}{CA402B}

\newcommand\vldbdoi{XX.XX/XXX.XX}
\newcommand\vldbpages{XXX-XXX}
\newcommand\vldbvolume{19}
\newcommand\vldbissue{1}
\newcommand\vldbyear{2026}
\newcommand\vldbauthors{\authors}
\newcommand\vldbtitle{\shorttitle} 
\newcommand\vldbavailabilityurl{URL_TO_YOUR_ARTIFACTS}
\newcommand\vldbpagestyle{plain} 
\newcommand{\ie}{{\itshape i.e.}, }%%

\begin{document}
\title{Towards Operationalizing Heterogeneous Data Discovery}

\author{Jin Wang}
\affiliation{%
  \institution{Megagon Labs}
}
\email{jin@megagon.ai}
\author{Yanlin Feng}
\affiliation{%
	\institution{Megagon Labs}
}
\email{yanlin@megagon.ai}
\author{Chen Shen}
\affiliation{%
	\institution{Megagon Labs}
}
\email{chen\_s@megagon.ai}
\author{Sajjadur Rahman}
\affiliation{%
	\institution{Megagon Labs}
}
\email{sajjadur@megagon.ai}
\author{Eser Kandogan}
\affiliation{%
	\institution{Megagon Labs}
}
\email{eser@megagon.ai}

\input{src/sec0-abstract.tex}

\maketitle

%%% do not modify the following VLDB block %%
%%% VLDB block start %%%
\pagestyle{\vldbpagestyle}
\begingroup\small\noindent\raggedright\textbf{PVLDB Reference Format:}\\
\vldbauthors. \vldbtitle. PVLDB, \vldbvolume(\vldbissue): \vldbpages, \vldbyear.\\
\href{https://doi.org/\vldbdoi}{doi:\vldbdoi}
\endgroup
\begingroup
\renewcommand\thefootnote{}\footnote{\noindent
This work is licensed under the Creative Commons BY-NC-ND 4.0 International License. Visit \url{https://creativecommons.org/licenses/by-nc-nd/4.0/} to view a copy of this license. For any use beyond those covered by this license, obtain permission by emailing \href{mailto:info@vldb.org}{info@vldb.org}. Copyright is held by the owner/author(s). Publication rights licensed to the VLDB Endowment. \\
\raggedright Proceedings of the VLDB Endowment, Vol. \vldbvolume, No. \vldbissue\ %
ISSN 2150-8097. \\
\href{https://doi.org/\vldbdoi}{doi:\vldbdoi} \\
}\addtocounter{footnote}{-1}\endgroup
%%% VLDB block end %%%

%%% do not modify the following VLDB block %%
%%% VLDB block start %%%
\ifdefempty{\vldbavailabilityurl}{}{
\vspace{.3cm}
\begingroup\small\noindent\raggedright\textbf{PVLDB Artifact Availability:}\\
The source code, data, and/or other artifacts have been made available at \url{\vldbavailabilityurl}.
\endgroup
}

\input{src/sec1-intro.tex}
\input{src/sec2-related.tex}

\input{src/sec3-techs.tex}

\input{src/sec4-challenge.tex}
\input{src/sec5-efforts.tex}
\input{src/sec6-conc.tex}

\bibliographystyle{ACM-Reference-Format}
\bibliography{ref/llmsys,ref/datalake,ref/other}

\end{document}

%% file: src/sec0-abstract.tex
\begin{abstract}
Querying and exploring massive collections of data sources, such as data lakes, has been an essential research topic in the database community.
Although many efforts have been paid in the field of data discovery and data integration in data lakes, they mainly focused on the scenario where the data lake consists of structured tables.
However, real-world enterprise data lakes are always more complicated, where there might be silos of multi-modal data sources with structured, semi-structured and unstructured data. 
In this paper, we envision an end-to-end system with declarative interface for querying and analyzing the multi-modal data lakes.
First of all, we come up with a set of multi-modal operators, which is a unified interface that extends the relational operations with AI-composed ones to express analytical workloads over data sources in various modalities.
In addition, we formally define the essential steps in the system, such as data discovery, query planning, query processing and results aggregation.
On the basis of it, we then pinpoint the research challenges and discuss potential opportunities in realizing and optimizing them with advanced techniques brought by Large Language Models. 
Finally, we demonstrate our preliminary attempts to address this problem and suggest the future plan for this research topic.
\end{abstract}

%% file: src/sec1-intro.tex
\section{Introduction}\label{sec-intro}
% abstract+intro: 1.5 to 2 pages

% para 1: overview of data lake and its importance
We have witnessed in the last decades a drastic growth in the number of open and shared datasets coming from governments, academic institutes, and private companies.
For instance, the size of US open data (\url{data.gov}) has reached a total of 335k datasets contributing to \$3 trillion of the US economy as of 2022~\cite{DBLP:conf/sigmod/Fan00M23}.
These massive collections of datasets, which are also known as \emph{data lakes}, open up new opportunities for technical innovation, economic growth, and social benefits~\cite{DBLP:conf/sigmod/Fan00M23,DBLP:journals/debu/MillerNZCPA18}.
To harness the inherited value from such huge data lakes, researchers from both industrial and academic fields have extensively explored various data-intensive tasks over data lakes, such as data discovery~\cite{DBLP:conf/sigmod/SarmaFGHLWXY12,DBLP:journals/pvldb/OtaMFS20,DBLP:journals/pvldb/FanWLZM23}, data integration~\cite{DBLP:conf/sigmod/YakoutGCC12,DBLP:journals/jdiq/LiLSWHT21,DBLP:journals/pacmmod/MiaoW23,DBLP:conf/sigmod/Wang0HK22}, information extraction~\cite{DBLP:journals/pvldb/LiFLMHLT19,DBLP:journals/pvldb/ThorneYSS0L21} and table question answering~\cite{DBLP:journals/pvldb/ZhangHFCDP24}.

\begin{figure}[h!t]
	\centering
	\includegraphics[width=0.5\textwidth]{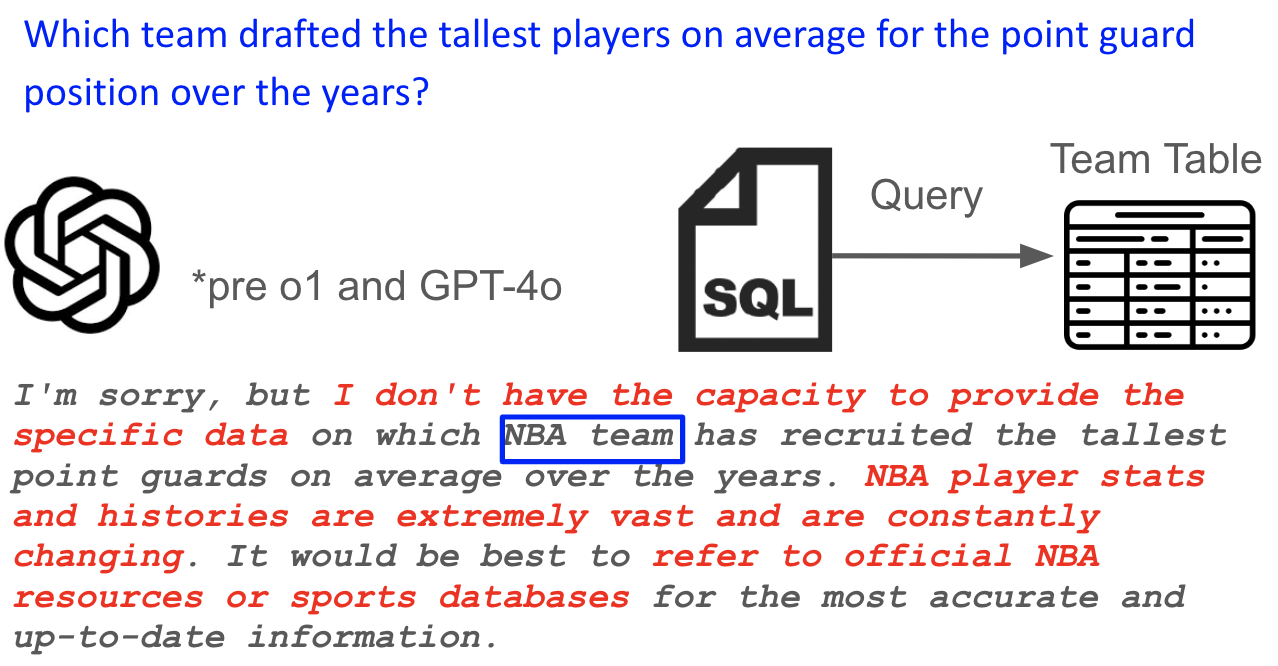}
	\caption{Example of Question Answering over Multi-modal Data.}
	\label{fig:motivation}
\end{figure}

% para 2: introduce the necessity of handling multi-modal data and the application scenarios.
Although many efforts have been paid in querying and analyzing data lakes, most existing works focused on the scenario where data collections are considered as structured data such as tables.
Nevertheless, modern enterprise data is often captured through datasets with various formats, such as tabular data with factoid information, graphs modeling symbolic knowledge, and documents with rich contextual information.
Besides, in the real-world enterprise data lakes, modern organizations (e.g., Merck, British Telecom, and the City of New York) organize data into thousands of sources that are managed by different teams and departments and often become silos of information~\cite{DBLP:journals/tkde/HaiKQJ23}.
Consequently, similar contents might be represented in \emph{multiple modalities} among different sources.
In order to answer a query in such scenario, the system might need to rely on different kinds of queries, e.g. relational operations, graph search and machine learning model inference, over data sources in multiple modalities.
A typical example could be found in Figure~\ref{fig:motivation}.

\begin{example}
	Figure~\ref{fig:motivation} shows an example of natural language question, which is considered as unstructured texts.
	When we attempted to directly answer it with prompts over LLMs, the responses indicated that these models could not effectively handle the question due to insufficient information about NBA teams.
	Meanwhile, if we have some table datasets about teams at hand, we can launch a SQL query to obtain the required information and then utilize them as part of the prompt over LLM.
	And such contexts could be helpful in getting the correct answer. 
	In this process, we can see that the information from the structured table definitely helps answer question in unstructured text.
\end{example}

% para 3: discuss related works like LOTUS and Symphony in more details about their limitations in handling above workload
There are some previous efforts about question answering on multi-modal data~\cite{DBLP:conf/cidr/ChenG0FM023,vldb25eleet} as well as defining semantic operators~\cite{DBLP:journals/corr/abs-2407-11418} by extending the relational ones to non-relational data formats.
However, these studies mainly regarded Language Models as the core of the solution rather than building an end-to-end system that provides a unified and declarative query interface.
Meanwhile, the vision introduced in this paper could support various kinds of queries, including structured queries, model inference as well as prompt over language models, on different kinds of data formats.
In other words, it would be a generalized framework that could cover the use cases of existing works.

% para 4: come up with our vision of pipeiline and list each component as a bullet with descriptions

In this paper, we envision an end-to-end system to enable a declarative framework for querying and analyzing multi-modal data lakes.
At query time, users can specify their intent with a unified query interface in high-level programming language and the system could respond to the query by first understanding the query intention, and then automatically generating optimized query plans for producing the results.
To be more specific, the system starts with a \emph{Query Interface} that aims at providing an interface for users to express their query intention. 
We envision a declarative query interface in high-level programming language consisting of primitive operators to enable expressing rich semantics. 
In many cases, the data sources or modalities, cannot be clearly identified in the user input.
To address this issue, the \emph{Data Discovery} process would help identify the potential modalities to answer the user query.
We discuss about the design principles for data discovery from multi-modal data lakes and provide some initial efforts in benchmarking the task.
As users express the query intent with a declarative interface with primitive operators, we then need a \emph{Query Planning and Optimization} component to transform the program into query plans and making necessary optimizations.
We propose several guidelines in defining the scope of query plans and discuss the challenges to find a optimal plan. 
We also come up with some initial ideas of plan optimization and illustrate them with running examples. 
Finally, since we might need to answer the query based on multiple data sources, the output should be obtained by aggregating results from all of them. 
We provide a proof of concept of such a process and illustrate the potential opportunities of improvement based on Large Language Models.

% para 5: the organization of rest of the paper
The rest of this paper is organized as following:
Section~\ref{sec-related} surveyed the related work.
Section~\ref{sec-tech} envisioned the overall workflow of the multi-modal data lake system.
Section~\ref{sec-co} raised the research challenges and discuss potential opportunities.
Section~\ref{sec-effort} introduced our initial efforts towards realizing this vision, especially in the aspect of benchmarking.
Finally Section~\ref{sec-conc} concluded the whole paper.

%% file: src/sec2-related.tex
\section{Related Work}\label{sec-related}
% 0.5 page
% aim at illustrating how previous works failed to cover the topics  discussed in this paper

\subsection{Data Lakes}\label{subsec-rdl}

Data discovery from data lakes has been a popular topic in the data management community for several decades.
Given a dataset or data science task, data discovery aims at identifying datasets from the data lakes that are relevant~\cite{DBLP:conf/sigmod/Fan00M23,DBLP:conf/sigmod/YakoutGCC12}.
For example, table unionable search~\cite{DBLP:journals/pvldb/NargesianZPM18,DBLP:journals/pvldb/FanWLZM23,DBLP:journals/pacmmod/KhatiwadaFSCGMR23} aims at returning the tables that are unionable with the given query table;
Joinable table search~\cite{DBLP:journals/pvldb/ZhuNPM16,DBLP:journals/pvldb/EsmailoghliQA22,DBLP:journals/pvldb/Dong0NEO23} is the problem of discovering tables that are joinable with the given table;
Domain discovery~\cite{DBLP:conf/sigmod/YakoutGCC12,DBLP:journals/pvldb/ArmbrustDPXZ0YM20} is the task to identify the potential topics for columns in the tables.
There are also several benchmarking efforts in data lake related tasks~\cite{DBLP:journals/corr/abs-2402-06282,DBLP:conf/guide-ai/FengRFCK24,DBLP:journals/pvldb/DengCCYCYSWLCJZJZWYWT24} such as discovery, table search and retrieval.
From the aspect of systems, \textsf{Delta Lake}~\cite{DBLP:journals/pvldb/ArmbrustDPXZ0YM20} proposed a lower-level storage for different data formats.
And \textsf{LakeHouse}~\cite{DBLP:conf/cidr/Zaharia0XA21} aimed at managing the data warehouse and data lakes in a unified manner.

Some recent studies also explored the scenario of multi-modal data lakes.
\textsf{CMDL}~\cite{DBLP:journals/pvldb/EltabakhKEA23} developed a model for joint data discovery from both texts and tables.
\textsf{Symphony}~\cite{DBLP:conf/cidr/ChenG0FM023} is a system for natural language question answering over multi-modal data lakes.
\textsf{Thetis}~\cite{DBLP:conf/edbt/ChristensenLLRM25} takes semantic knowledge into consideration for data discovery processes.
These works do not have a formal definition of operators and query plans as our introduced framework did. 

\subsection{Data Oriented Compound AI Systems}\label{subsec-raisys}

With the rapid growth of LLMs, compound-AI system has become an important concept in adopting LLM into various tasks.
It consists of interacting components, which are also known as agents, that could finish complex tasks in a collaborative manner~\cite{cais}.
Recently, there is a large amount of studies focusing on the data oriented Compound AI systems.
\textsf{ELEET}~\cite{vldb25eleet} jointly queried table and text data by devising new pre-training objectives for Transformer based language models.
\textsf{Palimpzest}~\cite{cidr25palimpzest} envisioned a solution for AI-powered data processing tasks.
\textsf{LOTUS}~\cite{DBLP:journals/corr/abs-2407-11418} defined a set of semantic operators with natural language literals to specify the predicates.
\textsf{DSPY}~\cite{DBLP:conf/iclr/KhattabSMZSVHSJ24} aimed at automatically optimizing the prompts to reduce the overall cost.
There are also other works aiming at utilizing LLM in query processing over both structured and unstructured data based on similar ideas~\cite{DBLP:conf/edbt/0002CP24,DBLP:conf/cidr/RajanRLRSP24,DBLP:conf/acl/ThorneYSS0H20,DBLP:conf/cidr/UrbanB24,cidr2025tag,DBLP:journals/pvldb/AroraYENHTR23}.

Although above studies have made some efforts in defining operators and planning techniques over tasks that potentially require datasets with multiple modalities, they mainly focused on the LLM-centric scenarios.
Meanwhile, our envisioned framework aims at providing formal definition for query processing over multi-modal data where LLM is not necessarily involved.
Actually, the envisioned framework could also support semantic operators where LLM based techniques could be seamlessly integrated into our framework.
We will illustrate this in the next sections.

%% file: src/sec3-techs.tex
\section{The Multi-modal Data Lake System}\label{sec-tech}
% sec3 + sec 4: 3 pages in total

\subsection{System Architecture}\label{subsec-system}

\begin{figure}[h!t]
	\centering
	\includegraphics[width=\columnwidth]{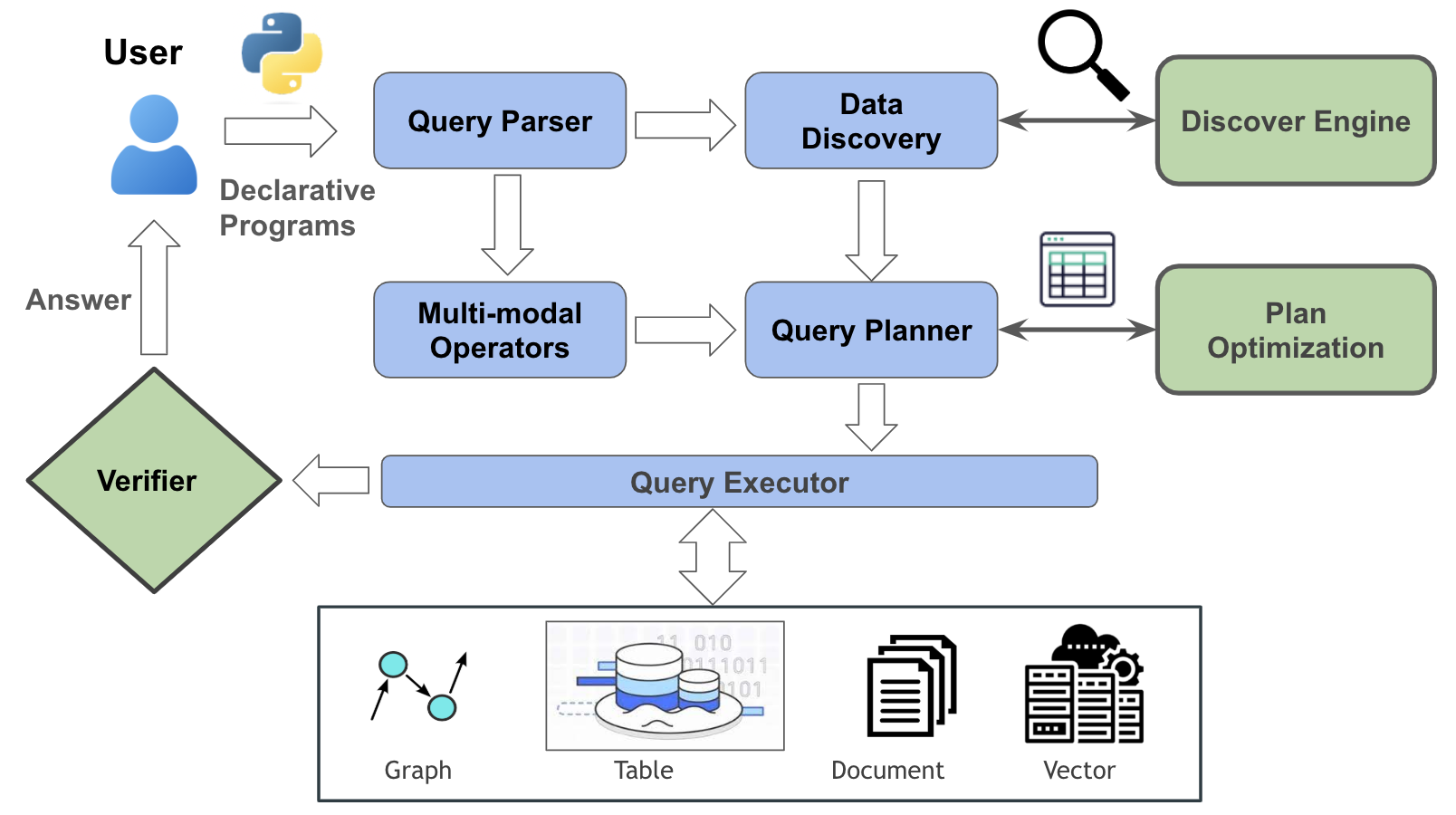}
	\caption{System Architecture}\vspace{-1em}
	\label{fig:overall}
\end{figure}

% \saj{Cuery parser, query executor}
The overall architecture of the multi-modal data lake system is shown in Figure~\ref{fig:overall}.
It could be considered as a query engine on top of multiple existing data storage or databases.
The multi-modal data is stored in each database, e.g. tables in PostgreSQL and texts in MongoDB, respectively.
And the proposed system would analyze the input query and then generate an overall query plan that might involve queries over different databases as well as other potentially required contexts.
In some cases, the required modality to answer the question is not clearly specified in the input query.
And we also need a data discovery module to address that issue.
Generally speaking, the system consists of the following key components:

\noindent\textbf{Declarative Query Interface}\hspace{.5em}
First of all, the system provides a declarative query interface to allow users to express their query intents.
We define the initial data model of the system as tables which contain either structured attributes or unstructured ones with free texts.
This could be implemented via popular libs such as Pandas~\footnote{https://pandas.pydata.org/}.
To express the query intent over multi-modal data in a unified manner, we provide a list of \emph{multi-modal operators} which could be considered as the primitive elements of a program.
More details of such parameters will be introduced in Section~\ref{subsec-mmops} later.
We implemented such multi-modal operators as functional APIs in a high-level programming language.
In the rest of this paper, we will use Python APIs as an example to illustrate the proposed techniques.
\smallskip

\noindent\textbf{Data Discovery}\hspace{.5em}
To generate a multi-modal query plan, it is essential to identify the targeted modality in each operator~\cite{DBLP:conf/guide-ai/FengRFCK24}.
Although the modality could be explicitly specified in the user input, in many cases users might not be able to recognize the required modalities due to the lack of knowledge about the data lake.
In this way, they need to rely on the data discovery module to identify the modalities that an operator would work on.
To reach this goal, we first need to build a mapping between contents in different data sources with the same or different modalities to identify the semantic relatedness.
Then given the query intent specified by users, the data discovery module would retrieve all data sources to identify the potentially relevant ones.
\smallskip

\noindent\textbf{Query Planning and Optimization}\hspace{.5em}
After identifying the data sources of each operator, the system then builds a multi-modal query plan that would be executed in each lower-level data storage/database system. 
The logical plan defines the dependency between operators as well as the overall workflow.
And the physical plan defines the behaviors of query execution in each data storage/database via translating the input programs into the query language in each specific system.
Meanwhile, it is essential to develop query optimization techniques to define the execution order of certain operators from different granularities.
We will make more discussion and provide some concrete examples about this in Section~\ref{subsec-mmops}.
\smallskip

\noindent\textbf{Verifier}\hspace{.5em} The final step of query processing is to obtain the results returned to users.
In the situation where the query plan has one root operator, this step is relatively easier to realize: we can directly regard the output of that operator as the output.
However, there might be situations where the results are based on the output of multiple operators.
In this case, the \emph{Verifier} component makes an aggregation over such outputs and generates a final answer.
There could be two straightforward solutions: a rule-based method uses some pre-defined heuristics to filter out irrelevant intermediate results and merge the relevant ones.
And a prompt-based method adds all results into a prompt template and asks LLMs to generate the output. 
Another potential solution could be involving human in the loop.
This could be realized by building an interactive user interface that allows users to identify the answers they want from the result or provide more instructions to refine the results.

\subsection{Multi-modal Operators and Query Plan}\label{subsec-mmops}

Next we would like to provide more insights about the design of multi-modal operators and the query plans consisting of them.
Similar to the previous work LOTUS~\cite{DBLP:journals/corr/abs-2407-11418}, we also define the operators by extending the relational model.
However, our proposed operators are different from those in LOTUS in the following aspects:
Firstly, we focus on generalizing the operators to different modalities in a high-level abstraction.
The definition is independent from how an operator is implemented, i.e. whether it is relational or semantic.
Secondly, the LOTUS operators utilize natural language expressions to specify the predicates.
While such expressions are flexible, they might not be able to illustrate the structured queries.
To address this issue, we allow users to program with the multi-modal operators by writing semi-structured predicates, e.g. JSON format, as the input argument.
In this way, it would enable expressing predicates for databases with structured query language and also provide flexibility for semantic queries over unstructured texts.
The details of multi-modal operators are introduced as follows:

\noindent\textbf{Selection}\hspace{.5em} It returns records from the database based on input predicates.
\smallskip

\noindent\textbf{Projection}\hspace{.5em} It returns a subset of attributes specified by the users over records from the database or output of another operator. 
\smallskip

\noindent\textbf{Join}\hspace{.5em} It performs a join operation between two collections of records with the same or different modalities. The predicates and types of join will be specified by users.
\smallskip

\noindent\textbf{Aggregation}\hspace{.5em} It performs different kinds of aggregation, e.g. min, max, average over the structured and semi-structured datasets. The semantics of the operators are equivalent with those defined in each modality.
\smallskip

\noindent\textbf{LookUp}\hspace{.5em} It aims at providing additional query semantics for text data, such as keyword search, similarity search and potential model inference over small and large language models.
\smallskip

\noindent\textbf{Ranking}\hspace{.5em} It provides a general ranking function for a collection of records. The criteria could either be specified by users or based on those defined in each database system.
\smallskip

\noindent\textbf{Conjunction}\hspace{.5em} It provides a general function to get the union, intersection and except results over two collections of records.

\begin{figure}[ht]
	\lstset
	{ %Formatting for code pieces
		language=Python,
		basicstyle=\footnotesize,
		numbers=left,
		stepnumber=1,
		keywordstyle=\color{blue},
		xleftmargin=0.5cm,
		showstringspaces=false,
		tabsize=1,
		breaklines=true,
		breakatwhitespace=false,
	}
	\begin{lstlisting} 
		lake = new DataLake(datalake_path);
		table_results = Selection(input = lake, src = "table", pred = {'attribute':'name', 'table': 'Team'})
		text_results = LookUp(input = table_results, src = "text", mode = "prompt",
		    pred = {'query': 'Which team drafted the tallest players on average for the point guard position over the years? '})
		Output(text_results)
	\end{lstlisting}  \vspace{-1em}
	\caption{Program: Natural Language Question Answering with additional information from Tables}
	\label{fig:prg1}
\end{figure}

\begin{example}
	The example program in Figure~\ref{fig:prg1} illustrates the way to answer the question shown in Figure~\ref{fig:motivation} with our envisioned system.
	Given an instance of a data lake, it starts from a \emph{Selection} operator on the that tries to find the names of all teams from the Team table, where the modalitiy is specified with the \texttt{src} input argument.
	Then such results would be utilized as the input for answering the question in text format with the help of a \emph{LookUp} operator. 
	It supports several modes: keyword search, similarity search and prompt over LLMs.  
	And the query input could be also specified in the predicates of the operator which is natural language question in this example.
	And the source of each operator is specified in the input argument of \emph{src}, respectively.
	And finally the \emph{Output} keyword specifies the output of the program.
\end{example}

In addition to above primitive operators, we also support various of User Defined Operators based on the specific needs in different modalities by providing a well defined template.
For example, in graph database there are usually path queries which do not belong to any primitive operators listed above.
In this case, users could just implement their own operators by inheriting the base class Operator and implementing a few key functions defined in it.
In addition, we also seamlessly integrate the data discovery module as part of the operators: if the targeted modality is not specified in the input argument of an operator, the data discovery process will be executed by default and no explicit function call is required.

\begin{figure}[ht]
	\lstset
	{ %Formatting for code pieces
		language=Python,
		basicstyle=\footnotesize,
		numbers=left,
		stepnumber=1,
		keywordstyle=\color{blue},
		xleftmargin=0.5cm,
		showstringspaces=false,
		tabsize=1,
		breaklines=true,
		breakatwhitespace=false,
	}
	\begin{lstlisting} 
		lake = new DataLake(datalake_path);
		join_results = Join(input = lake, pred = {'cond': ' find NBA games in the same location', 'cond_type': 'natural language', filter:{'year': '2023'}})
		project_results = Projection(input = join_results, pred = {'columns':'['team','score']'})
		Output(project_results)
	\end{lstlisting} \vspace{-1em}
	\caption{Program: Query without specifying the source.}
	\label{fig:prg2}
\end{figure}
\begin{example}
	Figure~\ref{fig:prg2} illustrates a program that aims at finding the participated teams and scores of the NBA games in 2023 in the same location. 
	However, the source of the query is not clear and thus it is not specified in the \emph{Join} and \emph{Projection} operators.
	To identify the source, the system will trigger a \emph{Discovery} operator implicitly when evaluating this program. 
	It will take the predicates of all operators and utilize the pre-defined data discovery operation to identify the source.
	In this program, the data source turns out to be table and then a SQL query having equivalent semantics with the program is executed on the database.
\end{example}

With the definition of such operators, we are then able to define the multi-modal query plans for input programs.
The logical plan could be regarded as a forest where the root node of each tree is considered as the output of the sub-plan.
Then the \emph{Verifier} is applied on top of all roots to obtain the final results.
And the physical plan consists of a query rewriting process to generate the corresponding query language, e.g. SQL, Cypher or functions in high-level programming languages, such as prompt over LLMs to work on each source, respectively.

\begin{example}
	We then consider a query equivalent to the natural language question ``which colleges in Massachusetts had player selections by teams that wins the champion in 1990s?'' over the multi-modal data lake.
	Suppose it requires the information of both graph and table data, then a potential query plan could be as follows:
	It first finds all teams that win the champion between year 1990 and 1999 from the table.
	Then it looks for the selected players of teams returned in step 1, and their college in the graph with a subgraph query and then apply the predicate that colleges should belong to Massachusetts. 
\end{example}

%% file: src/sec4-challenge.tex
\section{Challenges and Opportunities}\label{sec-co}

The envisioned system aims to build a unified query engine to support various workloads on multi-modal data with declarative query interface and effective cross-modal query planning techniques.
In this section, we will discuss the main next steps to realize this vision, including open research questions and associated challenges. \smallskip

\noindent\textbf{Unified Data Model.}\hspace{.5em} In the multi-modal data lake, one fundamental operation is to identify all relevant data sources for a given input query.
To this end, it is essential to propose a unified data model that could represent different modalities within one framework and build the cross-modal mapping between relevant records. 
The initial idea that formulates everything into a table with different kinds of attributes would fail to capture such kind of cross-modal information. 
Besides, additional efforts should also be paid to create structures over unstructured data, e.g. text and image, to fit the unified data model. 
Some initial efforts towards this goal is made in the previous work DataSpread~\cite{DBLP:journals/pvldb/BendreSZZCP15}, which aims to support interactions over heterogenous data types such numeric and textual values, mathematical formulae, and charts in Spreadsheets. 
However, the proposed data model still centered the table data.
% focuses on unifying spreadsheet interactions with in-database computation while guaranteeing positional correctness of a spreadsheet entry.
Finally, it is worthy investigating how to address entity relationship resolution so as to bride the semantic gap between data schema and values.
While such issues could be realized by the mechanisms of primary/foreign keys in relational databases, it would be rather challenging in the multi-modal scenario.
\smallskip

\noindent\textbf{Multi-modal Data Discovery}\hspace{.5em} In many situations, it is difficult for users to identify the targeted modalities of their queries. 
Thus a multi-modal data discovery component is necessary to fill this gap.
On the basis of the unified data model, there should be a data discovery solution over multi-modal data.
The CMDL framework~\cite{DBLP:journals/pvldb/EltabakhKEA23} proposed a customized method for the scenario with texts and tables, while our initial solution~\cite{DBLP:conf/guide-ai/FengRFCK24} employed prompt engineering to decide the involved modalities of a given question.
However, they fail to provide a general way to map data from different modalities to each other.
One reasonable solution to this problem is to first train a cross-modal representation learning model that can jointly model each modality into a unified vector space.
Next it generates an intermediate representation of each data source and sets a threshold for the relevance.
Then given the user input, it would compute the relevance scores based on the intermediate representation and return ones with higher score than the threshold as output.
Nevertheless, it would be rather challenging to obtain high-quality annotations for training such a model.
More efforts in the aspect of generating annotated data and devising training objective would be required here.
\smallskip

\noindent\textbf{Effective Query Optimization}\hspace{.5em} In the running example shown in Example 3.3, we provided one feasible query plan for the program.
However, there is still room for improving the performance.
Since we only needs the results of colleges in Massachusetts, if there are too many records about players from other states, the overhead of graph queries will be rather high. 
Then if the predicate is pushed to the query in table modality, such irrelevant results could be filtered and the performance could be significantly improved.

Now there are some early discussions about supporting such queries with the help of language models~\cite{DBLP:conf/edbt/0002CP24,DBLP:conf/cidr/UrbanB24,DBLP:conf/naacl/LiuXTSYL24,DBLP:journals/corr/abs-2408-00884}, but no concrete system level solution is provided yet.
Compared with traditional query optimization techniques within a relational database, there are more open problems to be explored in the scenario of multi-modal data lakes. Some examples include: 
How to automatically decide the execution order of queries in different modalities for a given program? 
Any opportunities of optimization via parallel execution of sub-plans in different modalities?
How to estimate the cost of queries over different modalities?
What kind of historical information should be collected to support future query optimization?
Could we apply predicates to another involved modality to reduce the cardinality of intermediate results?
\smallskip

\noindent\textbf{Handling Mixed Workload}\hspace{.5em} In addition to above query optimization issues, there are also potential challenges in handling workload with both relational query over different modalities and model inference.
As shown in recent studies~\cite{DBLP:journals/pacmmod/JoT24,DBLP:journals/corr/abs-2403-05821}, model inference or even prompt over LLM could be expressed in SQL-like query languages in the format of UDF.
In the background of our multi-modal operator, model inference could serve as the predicates of operators as well as the physical implementation of some operators.
For example, the LookUp operator could be implemented with reasoning over LLM or RAG to find the relevant information;
the aggregation operator over texts could be corresponding to the summarization task in NLP.
How to handle relational workload with such operations could be an important research topic. 
\smallskip

\noindent\textbf{Usability Enhancement}\hspace{.5em}
Another area to be explored is to enhance the usability of the system.
We have provided a set of functional APIs as the programming model for the system. 
Nowadays with the rapid growth of Large Language Models, providing the natural language query interface, such as text-to-SQL, become a hot topic again.
And how to build a unified natural language query interface over datasets with different modalities so as to make the proposed system more friendly to users, could be an interesting topic.
In addition, in the final output step, there could be results from different modalities.
It is necessary to provide a graphic user interface similar with the previous work for data discovery~\cite{icde25blend,DBLP:conf/icde/GongZGF23} and data lake exploration~\cite{DBLP:journals/pvldb/OuelletteSNBZPM21} to enable interactive exploration over the query results.
\smallskip

\noindent\textbf{Potential Enhancement with LLMs}\hspace{.5em} 
Finally, we can see that there are many opportunities to involve LLM into this multi-modal data lake system to improve the performance.
For instance, prompt over LLM could be considered as one potential solution for data discovery and result aggregation.
And it could also be utilized to implement some operators which can then support several NLP and data analysis tasks.
Meanwhile, there would be also some challenges to be resolved, such as defining the optimization goal of mixed workload and designing the prompt templates to realize a specific system component.
In addition, it is also worthy exploring how to build an agentic system of the proposed functionalities so as to seamlessly integrate the proposed techniques into a compound AI system, which could enhance the usability of the envisioned efforts in the context of LLMs.

% \red{TODO: would be good to introduce agentic system or agent-based query execution somewhere.}

%% file: src/sec5-efforts.tex
\section{Current Progress}\label{sec-effort}

% summarize the efforts in CMD bench and Simone's work to show our efforts in benchmarking towards this vision, 0.5 to 0.75 pages
In this section, we report our progress towards this vision.
Our initial efforts mainly focused on designing benchmarks for evaluating different components introduced in Section~\ref{fig:overall}. 

\subsection{Scope and Setting}  
\begin{figure}[!htb] 
  \centering
  \includegraphics[width=0.9\columnwidth]{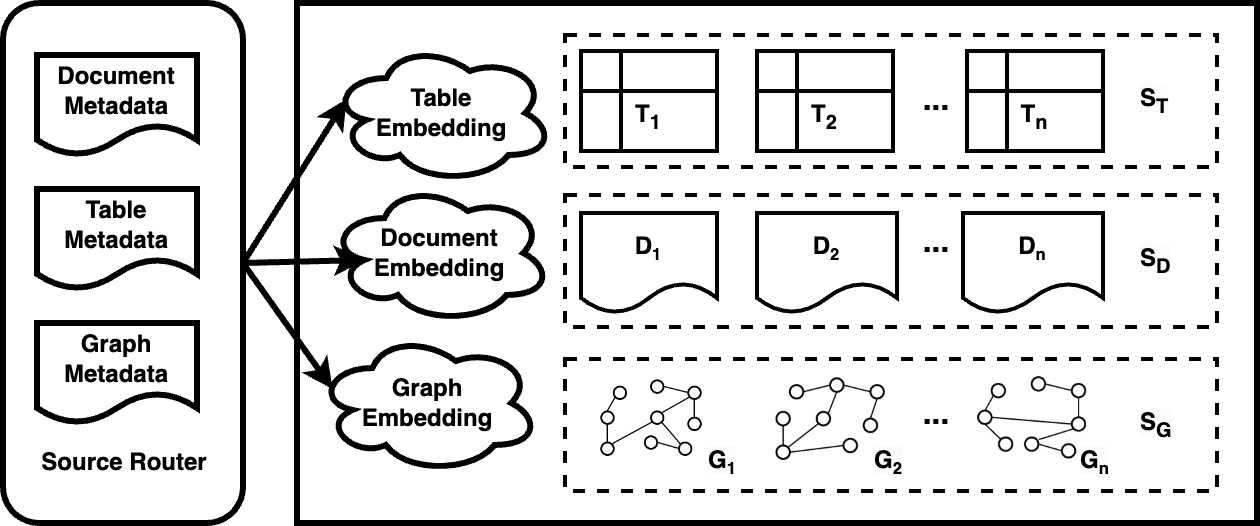}
  \caption{A simplified representation of the scope and setting with only one data source corresponding to basketball ($DS_b$). The tabular source ($S_T$) have statistics about players and teams, the graph source ($S_G$) contains symbolic knowledge and relationship among concepts, and the document source ($S_D$) has additional contextual information.}
  \label{fig:scope} 
  \vspace{-10pt}
  \Description{MMD lake.}
\end{figure}

We consider a streamlined setting (Figure~\ref{fig:scope}) in which, given a user query, the goal is to identify relevant data sources that can provide an answer, \ie source discovery. 
Once candidate sources are identified, the final response is generated through retrieval or structured queries. 
To assess both coarse- and fine-grained source discovery and task execution, we adapt existing datasets and benchmarks from open-domain question answering, complex reasoning tasks, and text-to-SQL.

 \vspace{-.5em}

\subsection{Evaluation: Data Discovery}
 \begin{figure}[!htb] 
	\vspace{-10pt}
	\centering
	\includegraphics[width=0.6\linewidth]{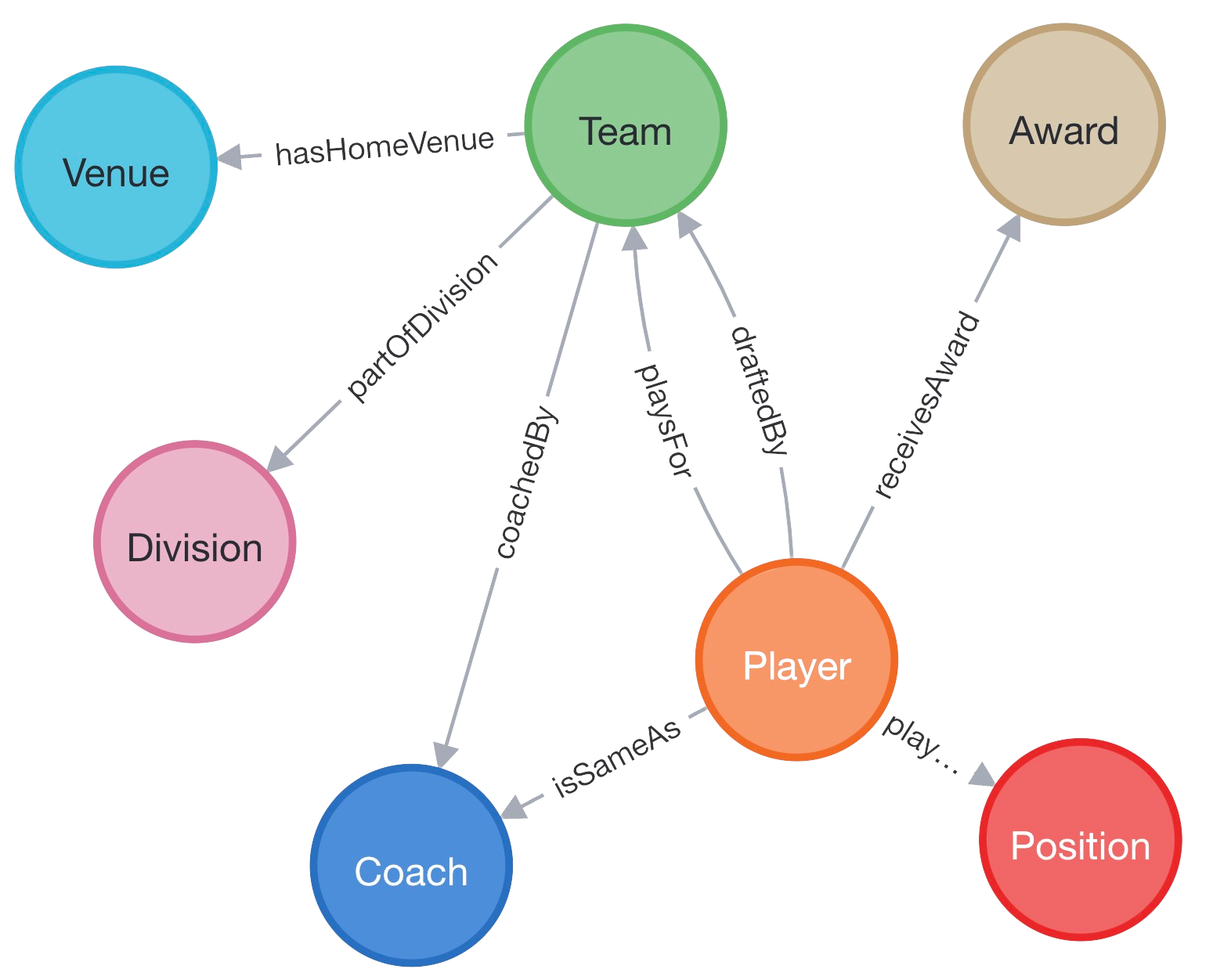}
	\caption{Schema of the NBA graph extracted from Wikidata.} 
\label{fig:graph_nba} 
\end{figure}

We constructed a comprehensive benchmark \emph{CMDBench}~\cite{DBLP:conf/guide-ai/FengRFCK24}, with data sources aligned with the setting in Figure~\ref{fig:scope}, focusing on the NBA (National Basketball Association) domain. 
This dataset integrates and adapts existing structured and unstructured sources to simulate the heterogeneous data scenario, including Wikipedia-extracted documents~\cite{Petroni2020KILTAB} and tables~\cite{zhongSeq2SQL2017}, and a knowledge graph derived from Wikidata~\cite{vrandevcic2014wikidata}.
An example of graph modality of this dataset is shown in Figure~\ref{fig:graph_nba}. 
To systematically evaluate unimodal discovery models, we design targeted discovery tasks using natural language questions adapted from well-established benchmarks in knowledge-intensive language tasks~\cite{Petroni2020KILTAB,zhongSeq2SQL2017,kwiatkowski2019natural,joshi2017triviaqa,cao2022kqa,yang2018hotpotqa}. 
Beyond task execution, we establish an additional evaluation objective centered on source discovery, developing benchmarks and baseline methods to assess how effectively data discovery enhances downstream task performance in complex AI systems. 
Finally, we present preliminary results from experiments with several discovery models adapted from the recently popularized \emph{LlamaIndex} framework~\cite{Liu_LlamaIndex_2022}, highlighting its potential in this setting.

% The goal of the benchmark is to evaluate coarse- and fine-grained discovery performance of respective models/agents. Source discovery evaluates whether, given a task, the appropriate coarse-grained discoverable element, \ie source(s), can be discovered.
% However, to understand the impact of data discovery efficacy on the downstream task, it is important to evaluate the end-to-end task performance. Our experiments highlight how data discovery performance varies with data modality, task complexity, and model design. In fact, we observed a $46\%$ drop in accuracy across all modalities even after employing the best-performing discovery models.

The benchmark is designed to assess both coarse- and fine-grained discovery performance across different models and agents. 
At the core of this evaluation is source discovery—determining whether, given a task, the relevant coarse-grained discoverable elements, \ie data sources, can be accurately identified. 
However, to fully understand the impact of discovery efficacy on downstream task performance, an end-to-end evaluation is essential. 
Our experiments reveal how discovery performance is influenced by data modality, task complexity, and model architecture. 
Notably, we observed a striking $46\%$ drop in accuracy across all modalities, even when leveraging the best-performing discovery models, underscoring the challenges inherent in effective data discovery.

\begin{figure}[!htb]
%\vspace{-0pt}
    \centering
    \includegraphics[width=\linewidth]{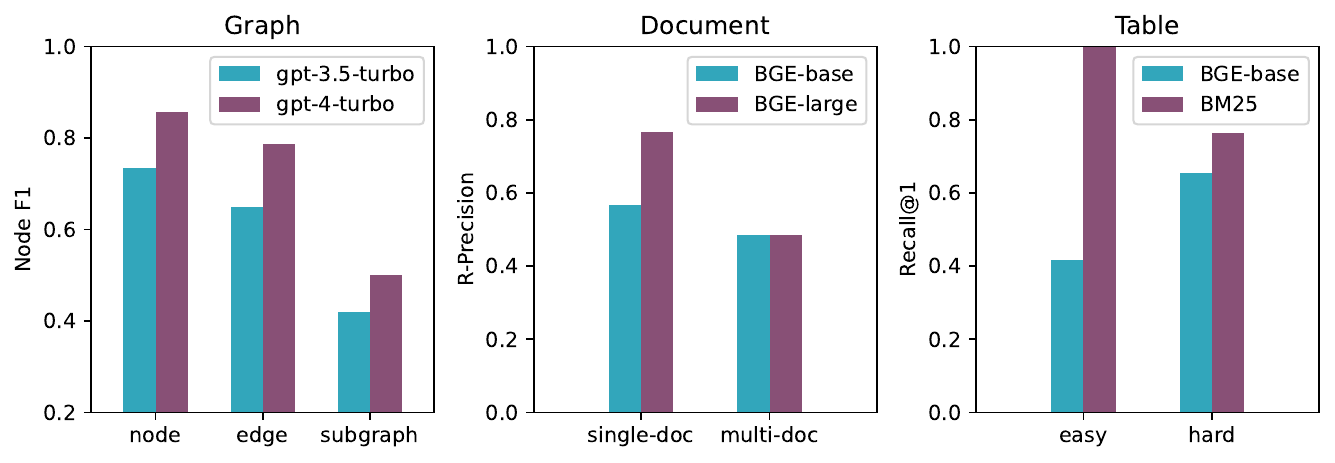}
    \caption{Source discovery on various question categories. }
        \vspace{-10pt}
    \label{fig:per_category}

\end{figure}

% Figure~\ref{fig:per_category} captures the impact of question difficulty on the source discovery performance for various modalities --- except for BGE-base embedding for Tables, the performance degrades with increasing question difficulty across all modalities. For tables, difficult questions tend to contain more contextual information, such as column headers, which may result in better table retrieval than easy questions, which tend to be terse. We show $R$-precision for documents due to the presence of multi-document HotpotQA~\cite{yang2018hotpotqa} questions. We report Node-$F1$ for graphs as we aim to measure the overlap between the retrieved sub-graph and ground truth. Since all of our tabular discovery tasks involved retrieving only one table, we report Recall$@1$.

Figure~\ref{fig:per_category} illustrates how question difficulty influences source discovery performance across various modalities. 
With the exception of the BGE-base embedding for tables, performance generally declines as question difficulty increases. 
In the case of tables, difficult questions often contain richer contextual details, such as column headers, which can enhance retrieval accuracy compared to simpler, more concise questions. 
We present $R$-precision for documents, given the multi-document nature of HotpotQA~\cite{yang2018hotpotqa} questions. 
For graph-based tasks, we report Node-$F1$, as it measures the overlap between the retrieved sub-graph and the ground truth. 
Since all our tabular discovery tasks involve retrieving a single table, we use Recall$@1$ for evaluation.

\subsection{Evaluation: Querying}

While robust benchmarks exist for evaluating document retrieval~\cite{Petroni2020KILTAB} and text-to-SQL~\cite{ DBLP:conf/emnlp/YuZYYWLMLYRZR18} in the LLM era, there is a notable gap in benchmarks for graph query generation from natural language. 
To address this, we introduce \emph{CypherBench}~\cite{DBLP:journals/corr/abs-2412-18702}, a dataset consisting of 11 property graphs derived from Wikidata. 
Each graph represents the \textit{complete} set of entities and relations from Wikidata, aligned with a domain-specific schema. 
Collectively, these graphs encompass 7 million entities, covering approximately 25\% of Wikipedia and 6\% of Wikidata. 
Additionally, we have developed over 10,000 natural language questions that span 12 distinct graph-matching patterns.

% \begin{table}[ht]
% % \begin{wraptable}{r}{0.55\textwidth}
% \scriptsize
% \centering
% \begin{tabular}{lccc}
% \toprule
% \textbf{Model} & \textbf{EX (\%)} & \textbf{PSJS (\%)} & \textbf{Exec. (\%)} \\
% \midrule
% \rowcolor[gray]{0.9}
% \multicolumn{4}{l}{\textit{Open-source LLMs (<10B)}} \vspace{2pt} \\
% \texttt{llama3.2-3b} & 11.20 & 17.33 & 86.46 \\
% \texttt{llama3.1-8b} & 18.82 & 30.98 & 90.67 \\
% \texttt{gemma2-9b} & 18.61 & 30.67 & 68.57 \\
% \midrule
% \rowcolor[gray]{0.9}
% \multicolumn{4}{l}{\textit{Open-source LLMs (10-100B)}}
% \vspace{2pt} \\
% \texttt{mixtral-8x7b} & 19.21 & 37.01 & 59.33 \\
% \texttt{qwen2.5-72b} & 41.87 & 56.39 & 86.84 \\
% \texttt{llama3.1-70b} & 38.84 & 54.79 & 92.25 \\
% \midrule
% \rowcolor[gray]{0.9}
% \multicolumn{4}{l}{\textit{Proprietary LLMs}}
% \vspace{2pt} \\
% \texttt{yi-large} & 33.82 & 47.21 & 83.52 \\
% \texttt{gemini1.5-flash-001} & 25.26 & 41.46 & 83.65 \\
% \texttt{gemini1.5-pro-001} & 39.95 & 57.70 & 86.03 \\
% \texttt{gpt-4o-mini-20240718} & 31.43 & 45.91 & 87.39 \\
% \texttt{gpt-4o-20240806} & 60.18 & 76.87 & 94.93 \\
% \texttt{claude3.5-sonnet-20240620} & \textbf{61.58} & \textbf{80.85} & \textbf{96.34} \\
% \bottomrule
% \end{tabular}
% \caption{Zero-shot execution accuracy (EX), provenance subgraph jaccard similarity (PSJS) and executable percentage (Exec.) on the CypherBench test set.}
% \label{tab:main_result}
% % \end{wraptable}
% \end{table}

\begin{table}[ht]
\scriptsize
\centering
\begin{tabular}{lc}
\toprule
\textbf{Model} & \textbf{EX (\%)} \\
\midrule
%\rowcolor[gray]{0.9}
\multicolumn{2}{l}{\textit{Open-source LLMs (<10B)}} \vspace{2pt} \\
\texttt{llama3.2-3b} & 11.20 \\
\texttt{llama3.1-8b} & 18.82 \\
\texttt{gemma2-9b} & 18.61 \\
\midrule
% \rowcolor[gray]{0.9}
\multicolumn{2}{l}{\textit{Open-source LLMs (10-100B)}}
\vspace{2pt} \\
\texttt{mixtral-8x7b} & 19.21 \\
\texttt{qwen2.5-72b} & 41.87 \\
\texttt{llama3.1-70b} & 38.84 \\
\midrule
% \rowcolor[gray]{0.9}
\multicolumn{2}{l}{\textit{Proprietary LLMs}}
\vspace{2pt} \\
\texttt{yi-large} & 33.82 \\
\texttt{gemini1.5-flash-001} & 25.26 \\
\texttt{gemini1.5-pro-001} & 39.95 \\
\texttt{gpt-4o-mini-20240718} & 31.43 \\
\texttt{gpt-4o-20240806} & 60.18 \\
\texttt{claude3.5-sonnet-20240620} & \textbf{61.58} \\
\bottomrule
\end{tabular}
\caption{Zero-shot execution accuracy (EX) on the CypherBench test set.}
\label{tab:main_result}
\end{table}\vspace{-1em}

% As shown in \autoref{tab:main_result}, the best-performing model \texttt{claude3.5-sonnet} achieves an execution accuracy of 61.58\% and a PSJS of 80.85\%, with \texttt{gpt-4o} performing slightly worse. The highest-performing open-source model reaches only 41.87\% execution accuracy, while smaller models in the <10B parameter range achieve less than 20\% execution accuracy. These results highlight the difficulty of CypherBench.

As shown in \autoref{tab:main_result}, the top-performing model, \texttt{claude3.5-sonnet}, achieves an execution accuracy of 61.58\%, with \texttt{gpt-4o} performing slightly behind. 
The highest-performing open-source model reaches only 41.87\% execution accuracy, while smaller models in the <10B parameter range fall below 20\% execution accuracy. 
These results underscore the challenging nature of CypherBench.

\iffalse

\fi

%% file: src/sec6-conc.tex
\section{Conclusion}\label{sec-conc}

This paper presents a vision for the data management community by demonstrating the potential opportunities of building a brand new system over multi-modal data lakes.
We came up with an initial design of novel query processing techniques by proposing a set of new multi-modal operators as well as introducing applications that could be expressed with them.
In addition, we also showcase a prototype built over an enterprise data lake as a proof of concept and provided some benchmarking tasks to evaluate the quality of data discovery and question answering over multi-modal data.
Based on such trials, we further raised some important technical challenges and came up with a discussion about many research directions to be further explored to realize such a new system.